\documentclass[%
 reprint,
 amsmath,amssymb,
 aps,
]{revtex4-1}
\usepackage{amsmath}
\usepackage{hyperref}
\usepackage{textcomp}
\usepackage{graphicx}
\usepackage{dcolumn}
\usepackage{bm}
\usepackage{amsthm, amssymb}
\usepackage{relsize,exscale}
\usepackage{amsmath}
    
    \usepackage{graphicx}
    \usepackage{dcolumn}
    \usepackage{bm}
    \usepackage{amsmath}
    \usepackage{exscale}

\begin{document}

\title{Quantum chaos and the spectrum of factoring}

\author{Jose Luis Rosales $^1$}
\email{Jose.Rosales@fi.upm.es}
\author{Vicente Mart\'{i}n$^1$}
\author{Samira Briongos$^2$}

\affiliation{%
 $^1$Center for Computational Simulation (Madrid) \\
 DLSIIS ETS Ingenieros Inform\'{a}ticos, Universidad Polit\'{e}cnica de Madrid,\\
Campus Montegancedo, E$28660$ Madrid, Spain.\\
 $^2$Integrated Systems Laboratory ETSI Telcomunicaciones, Universidad Polit\'{e}cnica de Madrid,\\
Avenida Complutense $30$, E$28040$ Madrid, Spain.\\}

\date{\today}

\begin{abstract}
There exists a Hamiltonian formulation of the factorisation problem which also needs the definition of a factorisation ensemble (a set to which factorable numbers, $N'=x'y'$, having the same trivial factorisation algorithmic complexity, belong).  For the primes therein, a function $E$, that may take only discrete values, should be the analogous of the energy from a confined system of charges in a magnetic trap. This is the quantum factoring simulator hypothesis connecting quantum mechanics with number theory. In this work, we report numerical evidence of the existence of this kind of discrete spectrum from the statistical analysis of the values of $E$ in a sample of random OpenSSL n-bits moduli (which may be taken as a part of the factorisation ensemble). Here, we show that the unfolded distance probability of these $E$'s fits to a {\it Gaussian Unitary Ensemble}, consistently as required, if they actually correspond to the quantum energy levels spacing of a magnetically confined system  that exhibits chaos. The confirmation of these predictions bears out the quantum simulator hypothesis and, thereby, it points to  the existence of a liaison between quantum mechanics and number theory.  Shor's polynomial time complexity of the quantum factorisation problem, from pure quantum simulation primitives, was obtained.
\vfil                             
{\it keywords:} {\bf \small Quantum Chaos; Quantum Simulation; Trapped Ions; Quantum Algorithms}
\normalsize                              

\end{abstract}


\maketitle

\section{Introduction}

Arithmetic and Quantum Mechanics share captivating similarities. For example, there is a typical probability distribution to measure some fixed distance between two prime numbers, similarly,  as is the  case, in  quantum physics, that there are different intensities for the observation of the transition between any two distant energy levels of the atom. Even more visual examples exist, for instance, Raman barcodes, emerging from nonlinear media quantum optics spectroscopy, are the counterpart of number theoretical congruence classes, being on the grounds of  optical readable code technologies. Thus, even though no confirmed connection between those two sciences exists to date, given the relevance of number theory in cybersecurity, discovering a possible deep connection between them will be of crucial interest. On the other hand, going to the fundamentals of analytical number theory, ideas about a possible  liaison between quantum mechanics and number theory emerged from  Hilbert and, independently, P\'{o}lya suggestions  (see Ref.~$1$) that Riemann's hypothesis (Ref.~$2$ and $3$) will  be trivially true if some Hermitian operator can be found such that its eigenvalues are the imaginary part of the zeroes of the complex Euler's function  $$\zeta(s)=\sum_{n=1}^{\infty} \frac{1}{n^s}=\prod_{p=prime}(1-p^{-s})^{-1}.\;$$  As a matter of fact, P\'{o}lya's hypothetical Hermitian operator could be assimilated to the Hamiltonian of some physical system and,  on these regards, the truth of Riemann's hypothesis implies that quantised energies exist that are the imaginary part of the zeroes of $\zeta(s)$. There is also numerical evidence that the statistical behavior  of these complex zeroes is related to the eigenvalues of large random Hermitian matrices (Refs.~$4$a and $4$b), an intriguing fact that also shares the statistics of the energy levels of magnetic quantum systems with anti-unitary symmetry breaking, i.e., a Gaussian Unitary Ensemble (a remarkable example of this is the Aharonov-Bohm billiard -- see Refs. $5$a, $5$b).

The most comprehensive program to implement these ideas, relating pure number theoretical conjectures with physics, was by Berry and Keating~(Refs. $6$ -- see also Refs.  $7$ and $8$). Nevertheless, these authors did not succeed to find a true bound Hamiltonian from which quantum discreteness would eventually emerge to cope  with the, also discrete, Riemann zeroes. In spite of this, since, modulus the truth of the hypothesis, the zeroes  would univocally determine the distribution of the primes, if the connection suggested by Hilbert and P\'{o}lya is correct, then,  there must also exist a quantum system whose energies universally give the primes themselves, and, since the primes are defined from Euclid's unique factorisation theorem, such a quantum system should determine a new and universal distribution of the possible prime factors of a number $N=xy$, product of two primes --clearly a finite and bound set because the possible lower factor will satisfy the simple constraint $x\leq\sqrt{N}$--. The number theoretical {\it energy function} has to be multiplicative with the primes $x,y$, and, given that the relevant quantity is the amount of primes not larger than $x$, i.e., the function $\pi[x]$, it has been  conjectured earlier by Rosales and Mart\'{i}n in Ref.$9$, that the  analogous to the energy of the physical counterpart of the factorisation problem should read as

\begin{equation}\label{E}
  E[x,N/x]=\pi[\sqrt{N}]^{-2}\pi[x]\pi[N/x].
\end{equation}

As a matter of fact, this arithmetic function may be considered, with the appropriate choice of canonical variables, as the Hamiltonian of an inverse harmonic oscillator of some physical system with confined trajectories.

\medskip

On the other hand, in the center of mass of every two ions in a Penning trap, the Hamiltonian of the magnetron degree of freedom exactly coincides with the inverse harmonic oscillator prescription in number theory and, wherefore,  Rosales and Mart\'{i}n in Ref.~$10$ suggested to model the factorisation problem on these physical grounds. Here we will generalize this model  upon adding a time  periodic electric quadru-polar perturbation. It is an important modification to the earlier autonomous proposal.  The key discovery of Ref.~$10$, that remains unaltered here, was that the integer discreteness of the number $N$  turns out to be a consequence of the proportionality of $N$  to the  quanta of the magnetic flux in the trap (trough the area of the largest magnetron orbit). This is the fundamental constraint which relates factoring with quantisation (Landau Levels). There are two additional remarks: firstly that, a time periodic perturbation yields to chaotic behaviors in the system's phase space  trajectories and,  secondly, that the presence of a magnetic field, which is required to make allowance for  the radial confinement  of the ions in the trap, will break the anti-unitary symmetry of the system. Therefore, these two considerations taken into account and given that there is no chaos in the quantum realm, according to Berry's criterium (Ref.~$5$ a),  there must exist a "{\it semiclassical, but non-classical, behaviour characteristic of systems whose classical motion exhibits chaos}", which is the program we intend to follow in this article. To such a degree, the available mathematical treatment will be that of modelling the quantum Hamiltonian in terms of a (complex) random matrix. Even though  the exact form of the matrix for the non-autonomous Penning trap is not specified, we should be able to apply the general theorems of the random matrix theory to provide the expected  unfolded level spacing  of the spectrum of the measurable energies.  Back in number theory one should anticipate  then that the probability distribution of  the function $E[x,N/x]$ (computed for the possible primes, $x\leq \sqrt{N}$), being the number theoretical counterpart of the physical energy of such a classically confined chaotic system,  should  follow the Gaussian Unitary Ensemble statistic (as from the theoretic  Aharonov-Bohm magnetic billiard case).

\medskip

The remaining of the paper is organized as follows: in section $II$ the Hamiltonian formulation of the factorisation problem along with the necessary concepts for the statistical treatment of the problem are provided. Section $III$ is devoted to the quantisation of the Hamiltonian introduced in the second section, which matches to that of a system of confined charges in a Penning trap. The spectrum of energies is calculated in the semiclassical approximation. Section $IV$ shows that, an experimentally realisable model of the quantum simulator, may be devised as the time average Hamiltonian of a confined system of charges (i.e., a Coulomb lattice) in a non-autonomous perturbed Penning trap.  Then, the existence of a new "radial breathing" degree of freedom, a consequence of the instability of the magnetron degree of freedom, is demonstrated. In the non autonomous model, the inverse harmonic oscillator Hamiltonian energies become those corresponding to the time average of a periodically perturbed Hamiltonian. In section $V$, given some factorable $N$, the probability distribution of the arithmetic function $E(x,N/x)$ is computed (for the probable prime factors not larger than $\sqrt{N}$). This distribution comes to be discrete, as  predicted from the spectrum of the measurable quantum  simulator energies. Given this discreteness, an inversion algorithm from this spectrum is equivalent to a factorisation algorithm with  polynomial complexity, i.e., it only requires resources scaling as a function of $\ln N$. Also in this section, we demonstrate that a Gaussian Unitary Ensemble probability distribution  fits to the number theoretical computations for the unfolded level spacing of the function $E$. This fact, indeed, represents a falsifiability test of the hypothesis of the quantum simulator of factoring. Our conclusions are summarised in section $VI$. Finally,  the proof  of the dynamic confinement of a Coulomb lattice having magnetron instabilities, in the presence of a non autonomous Penning trap, is found in the appendix.

\section{Hamiltonian formulation of factoring} There are many composed integers $N'$, such that $\pi[\sqrt{N'}]=\pi[\sqrt{N}]\equiv j$. It is convenient then to define the {\it Factorisation Ensemble}  $\;\mathfrak{F}(j)$  as the set of primes numbers, say $x_k,y_l$ whose products give numbers $N_{kl}$ with this property:

\begin{equation}\label{FactorisationEnsemble}
  \mathfrak{F}(j)=\{x_k,y_l,\; \text{primes}\; |\;  N_{kl}=x_k y_l, \;\; with \; j=\pi[\sqrt{N_{kl}}]  \}.
\end{equation}

The cardinal of this set is the amount of the different number theoretical energies $E[x,y]$  in the ensemble
 $\sharp\{\mathfrak{F}(j)\}\sim\sqrt{N}(\ln\ln\sqrt{N}+B),\;$ where $B$ is Meissel-Mertens constant (see Ref.$9$). Since this quantity is larger than the trivial algorithm complexity of factoring, $E$ is approximately degenerate, i.e., many $N'\in\mathfrak{F}(j)$ have almost the same energy. This prediction was  previously confirmed in Ref.~$10$.

\medskip

Asymptotically, the prime number theorem states that $\pi[x]\sim x/\ln x$, writing $h=\sqrt{N}$, we get
\begin{equation}\label{Asymtotic_Energy_Approximation}
E\sim 1+(\frac{\ln(x/h)}{\ln h})^2.
\end{equation}
Let us now compute the probability $\mathbf{P}_{E}$ for the energy function  defined in the factorisation ensemble.  First, if for each of the primes $x_k\in \mathfrak{F}(j)$,  the probability of being a factor of a given  $N$  is given by a function $\mathbf{P} (x)$, one has

\begin{equation}\label{Pnormalisation}
   1= \sum_{p =\;prime}\mathbf{P}(p)\equiv {\int}_{2}^h \mathbf{P}(x)\mathbf{D}\pi(x),
\end{equation}
where one uses a Lebesgue measure integration and  the sum is taken over the primes less than or equal to $h$. The Lebesgue integral runs over all the real numbers and, in order to compute it, we can take the approximation from the prime number theorem formulated for the density of the primes, i.e., $\mathbf{D}\pi(x)\simeq dx/\ln x$.  It gives $\mathbf{P}(x)=\ln x/h$. Moreover, since per each factorisation there is univocally a single $E$ function, we infer the existence of  the new Lebesgue measure  ${\mathbf{D} \displaystyle \mathbf{E}}=|\partial_x E|\frac{d\pi(x(E))}{dE}$, i.e.,

\begin{equation}
    \int_{2}^h \mathbf{P}(x)\mathbf{D}\pi(x)={\int}_{1}^{E[2]}\mathbf{P}_E\mathbf{D} \displaystyle \mathbf{E},
\end{equation}

to such a degree one asymptotically obtains, using Equation~(\ref{Asymtotic_Energy_Approximation}),
\begin{equation}\label{PE}
    \mathbf{P}_E\sim \frac{1}{2}\frac{(\ln h)^{2}}{\sqrt{E-1}}.
\end{equation}
Recall also that, number theoretically, there are two positive independent arithmetic functions, depending on $\pi[x]$, $\pi[y]$ and $j$, that can be built, namely
\begin{equation}
p[x,y]=\frac{1}{2}(\pi[y]-\pi[x])/j,\;\; q[x,y]=\frac{1}{2}(\pi[x]+\pi[y])/j,\;
\end{equation}
that suggests to write
\begin{equation}\label{Energies}
  H(p,q)=p^2-q^2
\end{equation}
which can be evaluated for every pair of primes $(x,y)\in \mathfrak{F}(j)$, i.e.,
\begin{equation}\label{H_E}
H(p,q)=-E
\end{equation}
with  $E=\pi[x]\pi[y]/j^2.\;$
Trivial solutions are $$q=\sqrt{E}\cosh (t+t_E),  \;\; \; and \;\;\;  p=\sqrt{E}\sinh (t+t_E)$$ where
$t$ should be considered as a quasi-continuous "time coordinate" and $t_E$ is a constant depending on $E$. Then, neglecting $\delta E/\delta t$, i.e., for $x=O(h)$, we get
$$
\dot{q}(t)=p(t)\equiv \partial_p H, \;\;   \dot{p}(t)=q(t) \equiv -\partial_q H,
$$
which means that, asymptotically, for large $N$, the arithmetic function $H(p,q)$ behaves exactly as expected for the Hamiltonian of a {\it negative energy}  inverse harmonic oscillator. Now, in order for the physical analogy to be fully consistent, the actual system that simulates the solutions of the factorisation problem should be confined, i.e., it has a bound set of possible classical trajectories in the phase space.  Let us now describe how a bound for the primes in $\mathfrak{F}(j)$ can be computationally built.  Given the finiteness of  the ensemble there will always exist a minimum bound for the lower factor of any $N_{kl}=x_k y_l\in\mathfrak{F}(j)$, say  $x_g$, which also belongs to this set. Specifically, let us define a "gauge parameter" $g$ as follows
$$x_g(k)= \lceil h^{2/3}(\ln h)^g-k\ln h\rceil,$$
where, $g$, which is a  real number, can be selected $g=O(1)$. Here, the integer $ k $ indicates that the limit must be taken for the prime numbers that, according to the prime number theorem, are separated from each other by a {\it unit of distance} of the order of $ \ln x_g \sim \ln h $, on average. For the arithmetic function $q$ it imposes
\begin{eqnarray}\label{AN1}
 q\leq q_g[x(k), N/x(k)],
\end{eqnarray}
which asymptotically scales as a function of the gauge $g$ and the integer $k$,
\begin{eqnarray}\label{AN}
 q_g(k)\sim  2/3 k(\ln h)^{-2g}h ^{-1/3}+ h^{1/3}(\ln h)^{-g}.
 \end{eqnarray}

\section{Stationary quantum states} Let us consider a confined system of ions (or electrons) characterised by some fundamental frequency $\omega_0$, a unit of mass $m$ and a charge $e$ in a hyperbolic Penning trap. Radial and axial confinement are driven by means of a static electric field and an axially oriented constant magnetic field.  In the ions (or electrons)  center of mass coordinate system the electric and magnetic forces are balanced  when the charged particles lay on exactly opposite radial positions near the center of symmetry of the trap, i.e., in the  electrostatic saddle point region. In this equilibrium state the  magnetron degree of freedom Hamiltonian becomes that of the factorisation problem, i.e., an inverse harmonic oscillator with negative energies. This implies  that, with such a simplified configuration of the trap, the balance of magnetic and electric forces should be unstable.

\medskip

In the quantum realm, nonetheless, there is no instability because we should extend the state  coordinates of each particle to include the spin coordinate $s_i$.  The state of the system is then a tensor product of  entangled   -parity preserved- states  of every pair of indistinguishable particles
\begin{widetext}
\begin{equation}\label{Entangled_State}
  \Psi[\{q_i^{(1)}\},\{q_i^{(2)}\};\{s_1^i\},\{s_2^i\}] = \prod_i\{ \psi (q_{i}^{(1)},\mathbf{s}_1^{i})\psi(q_{i}^{(2)},\mathbf{s}_2^{i})\pm \psi(q_{i}^{(2)},\mathbf{s}_1^{i})\psi(q_{i}^{(1)},\mathbf{s}_2^{i})\},
\end{equation}
\end{widetext}
where  the product is extended to every particle pair and the $+$ or, $-$ sign corresponds to whether the ions are either bosons or fermions. Then $q_{i}=|q_{i}^{(2)}-q_{i}^{(1)}|$, the relative distance between each pair, becomes a c-number of the individual state of each entangled pair. In what follows we will consider that the full state of the system consists of the tensor factorisation  in Equation~\ref{Entangled_State}.  The system's initial state is described only in terms of the relative distance between the ions and the parity entangled spin state $\chi_{i}(s_i^1,s_i^2)=\psi_1 (\mathbf{s}_1^{i})\psi_2(\mathbf{s}_2^{i})\pm \psi_2(\mathbf{s}_1^{i})\psi_1(\mathbf{s}_2^{i})$ of each of the interacting pairs, i.e.,
$$ \Psi_{\mathfrak{F}(j)}[\{q_i\};\{\mathbf{s}_i\},0]=\prod_i \chi_{i}(s_i^1,s_i^2)\sum_{kl}^{D[N]}\{\sum_{g}a_{kl}^{(g)}\phi^{kl}_{g}(q_i)\}.\;$$
where $\phi^{kl}_{g}(q_i)$ are the eigenfunctions of the Hamiltonian of the magnetron degree of freedom of each pair in the trap (which may depend on the boundary conditions, denoted here by the gauge parameter $g$). On the other hand, $D[N]=\sharp\{\mathfrak{F}(j)\}$ maps the cardinality of the factorisation ensemble with that from the Hilbert space of the physical system.

\medskip

With this picture in mind, there will be a probability $|a_{kl}|^2$ to measure the magnetron energy eigenvalue,  $\tilde{E}_{kl}$ say, proportional to the arithmetic function $E[x_k,y_l]$. This  corresponds to the factorisation of the number $N_{kl}=x_ky_l\in \mathfrak{F}(j)$. At time $t=0$, the quantum state of the system is exactly solved once we  determine the complex amplitudes $a_{kl}$.

\medskip
In order to get the quantum theory an additional theoretical abstraction is required: we declare that the canonical arithmetic functions $p$ and $q$  are quantum operators acting on the state of the confined physical system.
\medskip

Let us now land into physics from number theory upon providing dimensionally measurable canonical coordinates from the known arithmetic functions:  $$p=\hat{p}/ \sqrt{\hbar\omega_0 m} ,\;\; q= \varrho\sqrt{m\omega_0/\hbar}\;\;\; and \;\;\; E= -2\tilde{E}/\hbar\omega_0 ,$$
the system then satisfies the energy constraint   $$\frac{\hat{p}^2}{2m}-\frac{m}{2}\omega_0^2 \varrho^2 =\tilde{E}.$$
This means that for the confined system, there is a Hamiltonian whose eigenvalues   $E_{kl}$  label the allowed physical states that the quantum factoring algorithm operates with.  If the system  corresponds to some confined set of particles, say, the state at  $t=0$   would be  $\Psi_{\mathfrak{F}(j)}(q,0)$  with the appropriate bound conditions, e.g.,  $\Psi_{\mathfrak{F}(j)}[q_g,0]=0,\;$  where  $q_g$  represents the size of the box where the system is confined.  The full state wave function is written as a series of all its quantum states labeled by $\{kl\}$, or
\begin{equation}\label{State}
\Psi_{\mathfrak{F}(j)}[\{q_i\},t]=\prod_i \chi_{i}(s_i^1,s_i^2)\sum_{kl}^{D[N]}\sum_{g} a_{kl}^{(g)} e^{-i E_{kl} t/\hbar}\phi^{kl}_{g}(q_i).
\end{equation}
The  simulator is programmed with the number N depending on the values of the wave function on the boundary. The spectrum of frequencies depending on $q_g$ is the Fourier transform of the  autocorrelation function
\begin{widetext}
\begin{eqnarray}\label{Spectrum}
\nonumber \mathfrak{E}(\omega;N)=\frac{1}{2\pi}{\int}_{-\infty}^{\infty}dt e^{-i\omega t}{\int}_{0}^{q_g}\{dq_i\}\Psi_{\mathfrak{F}(j)}[\{q_i\},0]^*\Psi_{\mathfrak{F}(j)}[\{q_i\},t].
\end{eqnarray}
\end{widetext}
As usual, the only possible output of the simulator should be its allowed frequencies $E_{k,g}/\hbar$ with probability $|a_{kl}|^2$  which are the expected outputs of the quantum algorithm of factoring. Boundary conditions  for $\psi(q_i,0)$,  for the radial wave function of each entangled spin state, read
$$\psi(q_i,0)=\frac{1}{\sqrt{q_g}} \; \text{if }\;\sqrt{\mathit{E}}\leq q_i\leq q_g(k).$$
The full Hamiltonian of the confined system of particles is $H[\{p_i\},\{q_i\}]=\sum_{i} p_i^2-q_i^2$. The transit to quantum mechanics comes from the usual substitution $p_i\rightarrow -i\partial_{q_i} $, which leads to the Schr\"{o}dinger equation of the simulator of factoring (hereafter, to simplify notation, we will drop the particle index $i$)
\begin{equation}\label{Schrodinger}
\partial_q^2 \phi(q)+q^2\phi(q)=E\phi(q),
\end{equation}
with the  proposed boundary conditions for $\phi(q)$. It leads univocally to the spectrum of energies. To solve this problem, let us develop the solution in the semiclassical regime. This method, as a difference with the exact one, given in Ref.~$10$, provides a physical meaning for the number theoretical --rather arbitrarily introduced--  gauge parameter $g$. The WKB wave functions are

\begin{equation}\label{WKB}
            \phi^{kl}_{g}(q)\sim p^{-1/2}\sin \big\{
\int pdq+\vartheta_{g}^{kl} \big\} ,\;\;\; \sqrt{E}\leq q \leq  q_g(k),
\end{equation}
where $\vartheta_{g}^{kl}$ is global a phase depending on the gauge $g$. Far from the turning point at  $q=\sqrt{E}\;,$ we take the approximation   $\; p\simeq q-\frac{1}{2 q} E,\;$  obtaining

\begin{eqnarray*}
   \phi^{kl}_{g}(q)\sim (q^2-E)^{-1/4}\sin\{ q^2/2-E/2\ln\frac{q}{q_g(k)}+\vartheta_{g}^{kl}\}.
\end{eqnarray*}

The condition $\phi^{kl}_{g}(q_g(k))=0$ leads to  $\vartheta_{g}^{kl}= \pi l- q_g(k)^2/2$  (for $l\in Z$) while the second condition  $\phi^{kl}_{g}(\sqrt{\mathit{E}})=0$ can  be satisfied if and only if $E$ is the solution of

\begin{equation}\label{Quantisation}
  2\pi l-E\ln\frac{\sqrt{E}}{q_g(k)}-q_g(k)^2+E=0.
\end{equation}

Now one develops $E\rightarrow 1+\varepsilon+O(\varepsilon^2)$, a method that is only possible when $2\pi l\sim q_g(k)^2$ implying that  the gauge $g$ is indeed a function of the mode $l$, i.e., Equation~(\ref{AN}) taken into account,
 \begin{equation}\label{Gauge}
 g(l)\sim 1+\frac{1}{\ln[(\ln h^2/2)^2]}\{1-2\pi l/Q_0^2\},
\end{equation}
where $Q_0=h^{1/3}(\ln h)^{-1}$. Finally, feeding these expressions into Equation~(\ref{Quantisation}) yields to the spectrum of energies
\begin{equation}\label{epsilon}
 E(k,l)\simeq 1+ \frac{4 k}{3(\ln h)^{3g(l)}\ln q_g(k)},
\end{equation}
that coincides with the solution obtained in Ref.~$10$. Equation~(\ref{epsilon}) should be compared with Equation~(\ref{Asymtotic_Energy_Approximation}).

In the semiclassical approximation the probability of the $k,l$ state becomes, for $q_g(k)\simeq h^{1/3}(\ln h)^{-g(l)}$,
\begin{equation}\label{probability-k-state}
  |a_{kl}|^2\sim |\partial_k E(k,l) |\mathbf{P}_{E(k,l)}\rightarrow k^{-1/2} ,
\end{equation}
and Equation~(\ref{PE}) was taken into account; then, up to an arbitray phase $\chi_l$,
\begin{equation}\label{Fourier_Coefficient}
a_{kl}\rightarrow k^{-1/4}\exp{\{-i\chi_l\}},
\end{equation}
which  is an important genuine quantum result: there exists a discrete universal spectrum of energies for the factorisation ensemble of any number $N$, a result that is indeed independently of its bit size. Moreover the result is consistent with the scalability of the quantum simulator, because these Fourier amplitudes do not depend on the initial configuration where the number  $N$  has been encoded, as it should be.  This feature demonstrates the consistency and the validity of the quantum factoring simulator model. Moreover, given that the energy is degenerate, depending on the  allowed  gauges in  the  $k$  labeled state, there could be in general many lines, labeled by the quantum number  $l,\;$   for the same state.

\section{Experimentally realizable quantum factoring simulator} 

In the saddle point region, where the particles become confined, there is an effective repulsive inverted harmonic oscillator potential, i.e.,denoting $\omega_0^2=\kappa/m$, $U(q)=-1/2\kappa \varrho^2$. In the center of mass system, the energy constraint becomes
$$\hat{H}(\hat{p},\varrho)\equiv\frac{1}{2}\{\frac{\hat{p}^2}{m}-\kappa \varrho^2 \}=\tilde{E}.$$
The Penning trap axial frequency is  $\omega_z=\sqrt{2}\omega_0$.

\medskip

An experimentally realizable model of the quantum simulator can be devised when the number of particles pairs increases. In that case the system should  be considered as a Coulomb lattice. Confinement is experimentally achieved through the presence of a stroboscopically driven periodic electric quadrupolar field perturbation of  strength $\lambda$  with a frequency $\omega_{\lambda}$.  In these practical situations a more convenient configuration of the Penning trap will be cylindric instead of hyperbolic and the effective equilibrium  of the electric and magnetic forces are achieved when the quadrupolar field frequency $\omega_{\lambda} $ is very close to that of the Penning trap magnetron degree of freedom . If the total number of particles in the confined  Coulomb lattice is $K$,  the time dependant Hamiltonian becomes,

\begin{equation}\label{Stroboscopic}
  \mathbf{H}_K(\{\hat{p}_i\},\{\varrho_i\},t)= \sum_{i=1}^{i=K/2} \{\hat{H}(\hat{p}_i,\varrho_i)+\lambda\kappa\varrho_i^2 \cos{2\omega_{\lambda} t}\}.
\end{equation}

In general an exact solution  of this problem can not be obtained and the trajectories are known to be chaotic. Therefore, in the quantum theory only the average time problem makes sense. Indeed, in the center of mass coordinate system, according to Feynman-Hellmann theorem, its time average Hamiltonian should be that of the inverse harmonic oscillator presented here (see Ref.~$11$).
\begin{equation}\label{timeaverage}
 \int_{-\pi/2\omega_{\lambda}}^{\pi/2\omega_{\lambda}}\frac{\omega_{\lambda}}{\pi} dt\langle \Psi(\varrho,t)|\mathbf{H}_K(\hat{p},\varrho,t)|\Psi(\varrho,t)\rangle\rightarrow \mathbf{E}.
\end{equation}
Where  $\mathbf{E}= (K/2)\tilde{E}+L\omega_{\lambda},\;$ is Floquet's quasi-energy  and  $L= \partial_{\omega_\lambda} \mathbf{E}\;$ is the conserved angular momentum. Along these lines, the Coulomb lattice rotates with the stroboscopic frequency $\omega_{\lambda}$. Therefore the energy of every parity preserved entangled ion pair can be defined as a time average
$$\tilde{E}=2/K(\mathbf{E}-L\omega_{\lambda})\rightarrow\langle \frac{\hat{p}^2}{2m}-m\frac{\omega_z^2}{4}\varrho^2\rangle.$$

\medskip
Moreover, as shown in the appendix, using the classical theory for the center of mass trajectories of every two ions, dynamic confinement is achieved upon assuming that the indistinguishable ions lay instantaneously in equilibrium positions.  Regarding to the collective motion of the Coulomb lattice, one obtains, analogously to the quantum Feynmann-Helmann Equation~\ref{timeaverage},  an average time Hamiltonian. Now, in order for this average energy function to coincide with that of the inverse harmonic oscillator above, the stroboscopic frequency $\omega_{\lambda}$  should match to that of the unstable Penning trap magnetron degree of freedom and, yet, this condition also originates a new degree of fredom for the radial coordinates of every two ions, as said, in exactly opposite positions (as required for the exact balance of electric an magnetic forces):  as shown in the appendix, the orbits  experience a time periodic expansion and contraction motion, with period $\pi/\omega_{\lambda}$. These are Mathieu's resonances of the system, and, indeed, they were seemingly observed by Affolter, Driscoll  and Anderegg in Penning trap confined Mg$^{++}$ ions experiments  in Ref.~$12$, where  the observed phenomenon was said to correspond to a characteristic {\it radial breathing} degree of freedom  for the periodically perturbed (collective) Coulomb lattice trajectories.

\medskip

To finalise this experimental proposal,  recall that in the semiclassical theory a new quantum number should be assigned to any periodic degree of freedom and, given that, in quantum simulator of factoring model, the integer $l$ arises precisely from the wave function conditions at the actual maximum and minimum turning points of the radial coordinate, we must necessarily conclude that the new quantum number $l$  corresponds to  this new classical {\it radial breathing} degree of freedom.

\section{ The spectrum of factoring}

A practical model of the factorisation ensemble is the set of all products of two primes  with the same number of bits,$\; n-1< \log_2 x'y'< n\;$ say. It represents an extension to the actual factorisation ensemble that, recall, refers to a single $\pi(\sqrt{N})$.

$$\mathfrak{F}(n)=\bigcup_{i=\pi[2^{n/2-1}]}^{\pi[ 2^{n/2}-1]} \mathfrak{F}(i).$$

\begin{figure}[hbtp]
\centering
\includegraphics[scale=0.4]{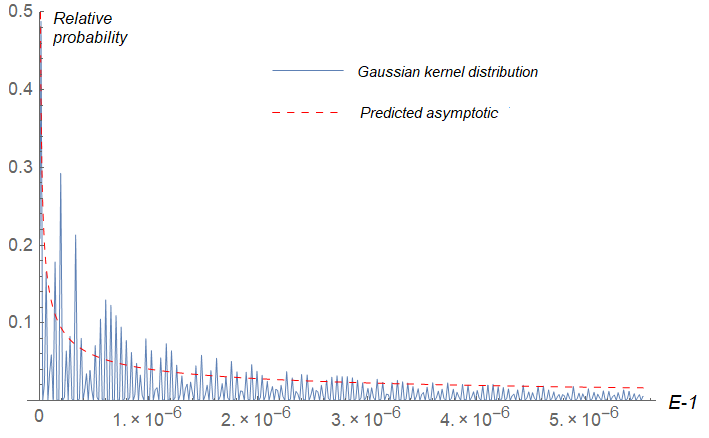}
\caption{Best fit Gaussian Kernel distribution calculated  for the  histogram of the factorisation function $(E[x',y']-1)$. The plot corresponds to the extended factorisation ensemble of a sample of $150,000$, $120-$bits, RSA public cryptographic key moduli. The predicted statistical behavior, Equation~(\ref{PE}), is the dashed red curve.}
\label{fig:Spectrum}
\end{figure}

As a result,  within the extended factorisation ensemble are the $n-$bits public moduli keys used in the RSA cryptography system.  The histogram of the function  $E[x',y']\;$  for a sample of these keys should fit to a universal discrete distribution of probabilities.  From the scalability of the spectrum to any size of the number $N=xy$  we are allowed to calculate $\mathfrak{F}(n)$ with arbitrary $n$,  e.g.,  $n=120$. Hence, a sample  of $150,000$ factorable $N\in\mathfrak{F}(120)$ RSA keys has been generated  using   OpenSSL. In order to perform a numerical experiment, we  generated $150,000$ values of  $E[x,y]\;$, using the aforementioned OpenSSL keys\footnote{To this aim, using the techniques in Ref.~$2$, we required the calculation of the  Riemann's series of $\pi[x]$, $\pi[y]$ and $\pi[j]$, with  $65,000$ zeta function zeroes, which is sufficiently accurate for these not very large bit size RSA keys.}. The Gaussian kernel  distribution histogram of the factorisation function is shown in Fig.~\ref{fig:Spectrum} which effectively displays the existence of a discrete set of favored values. Many $E$'s became apparently avoided while other are statistically amplified. The histogram represents the spectrum of factoring, confirming the expectations of the quantum theory for a system that classically exhibits chaos, as in the case of the confined ensemble of confined particles in a magnetic field.

\subsection{Polynomial complexity of the inversion algorithm}
As shown, for some low bit size RSA moduli, say $n'=120$, the discrete spectrum \[\mathfrak{E}[n'](E)=\sum_{kl}\delta(E-E_{kl}^{n'})|a_{kl}|^2,\] may be computed, to any desired  exactitude, upon adding  a finite number of zeroes in the Riemann series of $\pi[x]$, etc. Provided with this, we are now allowed to determine, with the same exactitude, other spectra corresponding to some much larger bit numbers, i.e., $N\sim 2^{n}$, $n\gg 120$. This comes from the fact that,  according to the factoring simulator model, Equations~\ref{epsilon} and \ref{Fourier_Coefficient}, there is a scalable spectrum of the universal simulator. This condition reads, explicitly
\begin{equation}\label{Scalability_of_the_Spectrum_of_the_energies}
E_{kl}^{n}=1+(n'/n)^4 (E_{kl}^{n'}-1).
\end{equation}
On these grounds, the spectrum encodes the universal probability distribution of the most likely factors of any number. Let us see how this remarkable prediction of the quantum theory may be used to find the more likely possible factors of any factorable number $N$, in principle, using polynomial resources. This algorithm  requires the inversion of the infinite Riemann's series of $\pi[x]$ and $\pi[N/x]$ in terms of the non trivial zeroes of $\zeta(s)$. The detailed techniques will be given elsewhere, however, let us advance here some the required tools and number theoretical methods.  To get $x=X[E_{kl},N]$ recall that, owing to Euclid's unique factorisation theorem,  for some known $N$,  the unique solution of the implicit constraint $$E_{kl}^n-E[x, N/x]=0$$ must be found.

\medskip

We now define the function  $$\varsigma_T(x)=1-  \sum_{\zeta(s_l)=0}^{T} \frac{R(x^{s_l})}{R(x)},\;\;$$
where $R(x)$ is Riemann's  approximation to  $\pi[x]=\lim_{T\rightarrow\infty}\varsigma_T(x)R(x)$. Then, up to some truncation order $T$ in the series of $\varsigma_T(x)$, a probable factor of $N$,  having probability $|a_{kl}|^2$, can be obtained if $x_0$ exists that minimise the constraints
\begin{equation}\label{InversioAlgorithm}
 (E_{kl}^n-E_T[x_0,N/x_0])^2\approx 0,
\end{equation}
where the notation $E_T(x,N/x)$ means that the replacements $\pi[x]\rightarrow \varsigma_T(x)R(x)$ etc., were used.  Then $x=\lim_{T\rightarrow\infty} x_0$.

\medskip

Notice that the function $\varsigma_T(x)$, owing to its definition as a series depending on the Riemann's zeros, suffers from large and rapid oscillations and, therefore, the constraints have many possible solutions.  In the end, the solutions of Equation~\ref{InversioAlgorithm} give  numerical approximations to the actual probable factors of $N$ (with the  given spectral probability $|a_{kl}|^2$). Yet, the exact factor $x$ can still be found.  One requires to feed $x_0$ into Coppersmith's algorithm  that computes an integer solution of a set of polynomial constraints of the kind
$$\mathbf{P_k}[z^k(x-\lceil x_0\rceil)]=0 \;\; mod \; N,$$
which, for the formally independent variable $z=x-\lceil x_0\rceil< 2h^{1/3}$, and $k\in \mathbf{N}$, form a set of problems that can be formally assimilated to that of finding the minimum reduced basis of a large lattice. Using the celebrated polynomial time LLL lattice basis reduction algorithm, the factor, $x=\lceil x_0\rceil +z$, will be obtained with  resources only scaling as $\ln N$ (see  Ref.~$14$ and Ref.~$15$).

\medskip

Provided with these  techniques, let us  theoretically estimate the best case factorisation algorithm complexity coming from the existence of  the spectrum of some $n$ bits size number $N\sim 2^n$, which, recall, is an scalable universal function of $\mathbf{K} \approx \mathbf{E}(\ln^4 h  /4 (E-1))$. Note first that there are \[\sharp\{\mathfrak{F}(j)\}/j\sim (\log_2 N)^2\log_2 \log_2 N\] constraints. Their solutions  provide all the possible approximations to the factors of $N$. On the other hand, if  $T$ becomes indefinitely large, the distance $|x-x_0|$ will necessarily be small, i.e., certainly not larger than $x^{1/3}$, say, which is the condition required for the applicability of Coppersmith's algorithm. In that case, the factor $x$ will be obtained in just $\log_2 N$ additional steps for every approximate solution of the constraints. This determines that the inversion algorithm obtains the factor $x$ in
\begin{equation}\label{Complexity_of_Factoring}
\Gamma \sim (\log_2 N)^3\log_2 \log_2 N
\end{equation}
steps, which exactly coincides  with the prescribed quantum  factoring algorithm complexity of Shor in Ref. $18$ for a quantum gate computer, as it should be. Notwithstanding with this encouraging result, recall that the best case corresponds to the exact summation of all the zeroes of $\zeta(s)$ in the series $$\varsigma(x)=\lim_{T\rightarrow\infty}\varsigma_T(x),$$ i.e., that the Riemann hypothesis must be true. In all practical purposes, though, the complexity  achievable with a classical computer that implements the inversion algorithm will strongly depend on the truncation order $T$. 

\subsection{Level spacings probability distribution}

As said in the introduction, the classical trajectories of the dynamically confined system will be chaotic. As a matter of fact, owing to the Von Neumann-Wigner theorem (Ref.$16$), the probability that two energy curves (depending on the strength $\lambda$) cross each other is extremely low, a phenomenon called level’s repulsion.  Considering that, in the  Coulomb lattice, there are  classical phase space trajectories having nearly the same semiclassical states,  one should conclude that only  the statistical distribution of the quantised energies can be studied.  This may correspond in number theory, we conjecture, to the fact that the value of the particular gauge $g(l)$  remains unknown. Hence, if two close  --{\it orbital}-- quantum numbers, say $l$ and $l'$, can be assigned to the same energy state corresponding to two nearly equally large {\it radial breathing} motions, one would expect that
 $$\Delta E(k,l)= E(k',l')-E(k,l) = s \hbar \omega ,\;$$ where $s$ is a random variable of non zero average. Thereby the quantum state can be described instead by the spectral statistic of the level spacing  $\wp(s)$. This procedure is, by construction, convenient for numerical studies.

\medskip

The action of  the Hamiltonian  on the state vector of such a chaotic or unpredictable system  can be replaced by the action of random matrices (see Refs.~$18$, $19$, and $20$). Therefore, level repulsion and randomness should become essential features of the energy distribution of the factoring simulator. Note that the presence of a magnetic field imposes that the system  has no time reversal invariance, which means that the matrices should have a complex Hermitian representation (see the net examples in Ref.~$5$ b). If the hypothesis of the simulator is correct, then, the expected distribution of the (unfolded, i.e. measured over the average) level spacing of the factorisation function $E$, in the ensemble of $n-$ bits RSA moduli, should be that of the Gaussian Unitary Ensemble which is given by the expression
\begin{equation}\label{GUE}
  \wp(\mathbf{s})=\frac{32}{\pi^2}\mathbf{s}^2\exp{( -\frac{4}{\pi}\mathbf{s}^2)},
\end{equation}

\medskip

\begin{figure}[hbtp]
\centering
\includegraphics[scale=0.55]{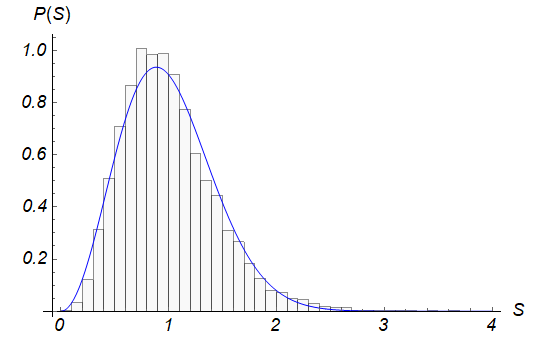}
\caption{ Histogram of the unfolded differences of the arithmetic function  $\{E[x_i,N_i/x_i]\}$ The continuous curve corresponds to the GUE Equation~(\ref{GUE}).}
\label{fig:GUE}
\end{figure}

We have  tested the validity of these physical ideas with numerical simulations regarding the distribution of the primes in $\mathfrak{F}(n)$. To do our analysis,  we computed  $E[x_k,y_l]$ for $500,000$ OpenSSL  $n-$bits RSA factorable moduli of the usual form $N_{kl}=x_k y_l\in \mathfrak{F}(n)$. Just for the sake of cross testing the results with the available table of primes in Mathematica$^\copyright\;$ we took $n=80$. Thereon, recalling the quantum  predicted energy function in Equation~(\ref{epsilon}), we  define the  $k$-index function
 $$\mathbf{k}(x_k,y_l)= \frac{1}{4}(E[x_k,y_l]-1) \{\ln [2^{n/2}]\}^4,$$
and we have taken into account that   $\ln h\simeq \ln [2^{n/2}]$  should be a good approximation. This arithmetic function is always $O(1)$ for any $n-$ bit RSA moduli and, according to the prediction of the quantum simulator, it should exhibit an universal probability density $ |a_{kl}|^2\sim \mathbf{k}^{-1/2}$  independently of the number of bits to which the extended factorisation ensemble pertains. Now, in order to calculate the unfolded level spacing, for the randomly selected $500,000$  samples in the extended factorisation ensemble, we must, first,  order $\left\{\mathbf{k}(x_k,y_l)\right\}$  from lowest to highest values to obtain an ordered set
  $$\left\{\mathbf{k}_i\right\}_{sampled}\rightarrow \left\{\mathbf{k}_i \right\}_{ordered}.$$
Moreover in order to avoid any possible bias in the definition of the closest energy level, we  computed the differences of almost consecutive values of the array of the ordered $k$-index function at the running $i$-th labeled position
 $$\Delta \mathbf{k}_i(\ell)=\frac{1}{\ell}(\mathbf{k}_{i+\lceil\ell/2\rceil}-\mathbf{k}_{i-\lceil\ell/2\rceil})$$
with the index $1\leq\ell\leq 6$ taken  as a random variable, i.e., $\ell=O(1)$, which is the only prescribed condition. This numerical procedure makes sense inasmuch as   we are trying to erase any kind of probabilistic bias originated from  the external program (in view of the fact that the pairs $(x_k, y_l)$ of the sampled OpenSSL generated primes were also randomly generated). Thereupon one proceeds to compute the average level spacing.  It requires to take into consideration  values in the array well beyond the actual level spacing  that we  are calculating at the position labeled by the index $i$. Numerically, we take some large  $L\gg Max[\ell]$ and define
 $$\langle\Delta \mathbf{k}_i\rangle_L \equiv \frac{1}{L} (\mathbf{k}_{i+\lceil L/2\rceil}-\mathbf{k}_{i-\lceil L/2\rceil}).$$  In the numerical experiment  $L=1000$ is taken (because it is much lower than the actual size of the sample, but is much larger than that considered for the nearby levels). The unfolded level spacing of the quantum index function at the running ordered position $i$ is then the random variable

\begin{equation}\label{LEVEL_SPACING}
  s_i(\ell)=\frac{\Delta\mathbf{k}_i(\ell)}{\langle\Delta \mathbf{k}_i\rangle_L}
\end{equation}

whose  normalised  histogram  is shown in Fig.~\ref{fig:GUE}. It fits exactly to the Gaussian Unitary Ensemble statistics, a result that is perfectly consistent with  the expected level repulsion of the quantum simulator with its associated number theoretical function $E$. The figure shows, for the primes in the extended factorisation ensemble of $80$ bits RSA moduli,  $\mathfrak{F}(80)$, the histogram of the unfolded differences of the arithmetic function  $\{E[x_i,N_i/x_i]\}$ calculated for a sample of $500,000$ moduli in this set. These primes were generated by the Unix standard cybersecurity package OpenSSL. In the quantum factoring simulator model those values  should be associated to the level spacing of the quantum factoring simulator energies with the prescribed level repulsion. This supports, by evidence, the predictions anticipated from the quantum theory on regards to the distribution of the primes in the extended factorisation ensemble.

\section{Conclusions} \par The hypothesis of the quantum simulation of the factorisation problem connects quantum mechanics and number theory. This is very analogous to Hilbert and P\'{o}lya conjecture to prove the Riemann's hypothesis related to the existence of a Hamiltonian system whose energy eigenvalues are the imaginary part of the non trivial Riemann's zeroes. The quantum simulator approach extends this connection to the primes. This proposal was previously introduced by Rosales and Mart\'{i}n in Ref.~$9$ and Ref.~$10$. Additionally, in this work, the semiclassical approximation of the energy levels probability distribution has been derived for the quantum states of the simulator. Extending the concept of the factorisation ensemble to cope with numerically computable RSA cryptographic moduli $N=xy$, i.e.,  to actual standard cryptograpic factorable  n-bits numbers, we have observed that the proposed  "energy factorisation function" statistical distribution is fully consistent with the predictions of the quantum model  (since  $E[x,y]\;$ correctly exhibits a {\it discrete} spectrum of probabilities). The asymptotic probability predicted dependence was also observed. This can be explained in the context of the quantum simulator model, but has no explanation whatsoever in the classical realm. To such a degree, then, the evidence provided here discovers an essential (i.e., quantum theoretical unavoidable) vulnerability of the RSA cryptographic system. On these   regards, we have developed an alternative and independent deduction of the polynomial time complexity of the quantum factorisation problem. This result, that comes from pure quantum simulation primitives, Equation~\ref{Complexity_of_Factoring}, requires the universality of the spectrum of the quantum simulator energies as well as the truth of the Riemann hypothesis.

\medskip

Finally, in this work, a crucial additional statistical test can be designed: if the exposed  quantum theory of factoring is correct, i.e., if the factorisation function $E$  corresponds to the actual energy of a magnetically confined set of charged particles, as suggested in Ref.$10$ and,  more explicitly described here, for the case of  a very special kind of stroboscopically perturbed Coulomb lattice system,  the probability distribution of the level spacing of the factorisation function  must be that of the Gaussian Unitary Ensemble and no other. This last test has also been numerically confirmed, a fact that affirmatively points out toward the existence  of a profound connection between quantum mechanics and number theory (since we have been able to confirm predictions that physics alone imposes on the distribution of the primes).

\medskip

\textbf{Acknowledgements} \par This work has been partially supported by the UPM contract number $P180021289$ and the Spanish Ministry of Economy and Competitiveness under contract RTC-$2016-5434-8$.
\vfil

\medskip
\textbf{References}\\
1 Montgomery H.L., ($1973$) Analytic number theory, in Proceedings of the Symposium on Pure Mathematics, St. Louis Univ., St. Louis, Mo., $1972$ (American Mathematical Society, Providence, R.I., $1973$), Vol. XXIV, pp. $181-193$. \\
2 Edwards, H. M. "Riemann's Zeta Function", New York: Dover, $2001$. \\
3 Riemann, B. $(1859)$", \"{U}ber die Anzahl der Primzahlen unter einer gegebenen G\"{o}$\beta$e" in Riemann, B. "Gesammelte Werke.", Teubner, Leibzig $(1892)$. \\
4(a)  Odlyzko A. M., ($1987$), "On the Distribution of Spacings Between Zeros of the Zeta Function", Mathematics of Computation Vol. 48, No. 177, pp. $273-308$ \\
4(b) Odlyzko A. M., ($1990$) "Primes, quantum chaos and computers", Number Theory, in Proc. Symp. National Research Council, Washington DC, $1990$, pp. 35–46. \\
5(a) Berry M. V. "The Bakerian Lecture, 1987: Quantum Chaology" ($1987$), Proc Roy Soc Lond Math Phys Sci  A,  $8(413)$, No. $1844$,  pp. $183-198$.\\
5(b)  Berry M. V. and  Robnik  M., ($1986$)  "Statistics of energy levels without time-reversal symmetry: Aharonov-Bohm chaotic billiards" ,J. Phys. A: Math. Gen. 19 pp. 649-668.\\
6 Berry M. V. and  Keating J.P., ($1999$) "The Riemann zeros and eigenvalue asymptotics", SIAM Rev. $41$, $236$.  \\
7 Schumayer D. and  Hutchinson D.A.W., ($2011$) "Physics of the Riemann hypothesis", Rev. Mod. Phys. $83$, $307$ \\
8 Sierra G. and  Townsend P. K., ($2008$), "Landau Levels and Riemann Zeros", Phys. Rev. Lett. $101$, $110201$. \\
9 Rosales, J.L. and  Mart\'{i}n, V. ($2016$) "Quantum Simulation of the factorization Problem", Phys. Rev. Lett. $117$, $200502$ \\
10 Rosales, J.L., Mart\'{i}n V. ($2018$)," Quantum simulation of the integer factorization problem: Bell states in a Penning trap", Phys. Rev. A 97, 032325.\\
11 Sambe, H., ($1973$),"Steady States and Quasienergies of a Quantum-Mechanical System in an Oscillating Field", Phys. Rev. A $7$, pp. $2203-2213$.\\
12  Affolter, M.,  Driscoll, C. F. and Anderegg, F. ($2014$), ``Space Charge Frequency Shifts of the Cyclotron Modes in Multi-Species Ion Plasmas'', J. of Amer. Soc.  Mass Spect. $26(2)$, DOI:10.1007/s13361-014-1030-9.\\
13  Brown, L.S. and Gabrielse, G. ($1986$ )  ``Geonium theory. Physics of a single electron or ion in a Penning trap,'' Rev. Mod. Phys. $58(1)$, pp. $233-311$.\\
14 Coppersmith, D., ($1997$), "Small solutions to polynomial equations, and low exponent RSA vulnerabilities", J. Crypt. $10(4)$, pp. $233-260$.\\
15. Lenstra, A. K.; Lenstra, H. W., Jr.; Lovász, L. ($1982$). "Factoring polynomials with rational coefficients". Mathematische Annalen. 261 (4): 515–534.\\
16 Shor, P.W.. ($1999$). "Polynomial-Time Algorithms for Prime Factorization and Discrete Logarithms on a Quantum Computer". SIAM Review $41 (2)$ pp. $303-332$.\\
17 von Neumann J. and Wigner J. ($1929$), "Uber das Verhalten von Eigenwerten bei adiabatischen Prozessen", Phys. Zeit., 30 , 467-470.\\
18 Dyson, F. J. ($1962$), "Statistical theory of the energy levels of complex systems I", J. Math. Phys., $3$, 140.\\
19 Porter, C. E. ($1965$), "Statistical theories of spectra: fluctuations", New York: Academic Press.\\
20 Haake, F. ($2001$), "Quantum Signatures of Chaos", Springer, Berlin, ISBN 3-540-67723-2, ($2$nd Edition). pp. $47-118$.\\
21 T. Hasegawa,  M. J. Jensen and  J. J. Bollinger, ($2005$) ``Stability of a Penning trap with a quadrupole rotating electric field'', Phys. Rev A $71$, $023406$.\\
22 Dubin   D. H. E. and  O'Neil T. M.  ($1999$) ``Trapped nonneutral plasmas, liquids, and crystals (the thermal equilibrium states)'', Rev. Mod. Phys. $71$, p.$87$.\\
23 Meirovitch, L. ($2003$) ``Methods of Analytical Mechanics''. McGraw-Hill Ed. ($1970$) reprinted by Dover Books, NY,  ISBN 0-486-43239-4, pp. $282-288$.\\

\appendix

\section{Dynamic confinement in Penning traps}

\medskip

Let us find the stable solution for the motion of two (ideally identically charged) clusters of ions in a Penning trap with a rotating wall. Radial symmetry is also taken into account. The case of many pairs of clusters to form a Coulomb lattice is straightforward using this symmetry.

In the Penning trap, the motion is decomposed into separated radial and axial ones. The system of particles in the trap is restricted to follow a harmonic oscillation in the $z-$ axis and a planar $(x,y)$ motion. For the $x,y$ plane of motion of two identical  charges at $x_1=-x_2=x$ and  $y_1=-y_2=y$, of total mass $M=2m$, the Lagrangian   is given in terms of the electrostatic quadrupole and the magnetic field frequencies of the trap
\begin{widetext}
\begin{eqnarray}
 \label{L1}
\nonumber \mathbf{L}_{2e}=\frac{1}{2} M(\dot{x}^2+\dot{y}^2)+\omega_z ^2 \frac{1}{4} M(x^2+y^2)+
\nonumber e\textbf{v}_1\cdot \textbf{A}(x,y)+e\textbf{v}_2 \cdot \textbf{A}(-x,-y)\\ \nonumber -\frac{e^2/2}{\sqrt{x^2+y^2}}-
\frac{M}{2}\omega_z^2 (x^2-y^2)\lambda\cos(2\omega_{\lambda}t)+M\omega_z^2 xy\lambda \sin(2\omega_{\lambda} t),
\end{eqnarray}
\end{widetext}
here $\textbf{A}(x,y)=-B y/2 \textbf{i}+B x/2\textbf{j}$ is the vector potential in the Johnson-Lippman gauge  and $\textbf{v}_1=
\dot{x}\textbf{i}+\dot{y}\textbf{j}$, $\textbf{v}_2=-\textbf{v}_1$. A periodic rotating quadrupolar electric potential wall was added. This term is  required for the adiabatic stability of the ions in the trap (see ~Ref.~$21$). The relative intensity of the rotating wall $\lambda$ will be determined from dynamic equilibrium considerations of the confined ensemble of ions in the trap. Hence,  close to dynamic equilibrium, statistically, the ions should occupy positions in the trap satisfying approximately, for their polar radius   $x^2+y^2\simeq a^2$, in terms of some constant distance to the center $a$,that  will be determined below using the dynamic equilibrium conditions. Moreover, one can write, denoting $\varrho=\sqrt{x^2+y^2},$ $$\frac{1}{\varrho} = \frac{1}{4a}(\frac{\varrho^2}{a^2}+3)+\dots,\; $$ then, for each of every two approximately identical charged density clumps near their equilibrium position, that is, disregarding higher order  terms, obtains the approximate quadratic Lagrangian
\begin{widetext}
\begin{eqnarray}
\label{L2}
 \nonumber \mathbf{L}_e\rightarrow\frac{1}{2} m(\dot{x}^2+\dot{y}^2)+\frac{1}{2} m(x^2+y^2)(\frac{\omega_z ^2}{2}+\frac{\beta}{a^3} ) +\frac{m}{2}\Omega(x\dot{y}-y\dot{x})\\ - \frac{1}{2}m\omega_z^2 (x^2-y^2)\lambda \cos(2\omega_{\lambda} t)+m\omega_z^2 xy\lambda \sin(2\omega_{\lambda}t),
\end{eqnarray}
\end{widetext}
where $\Omega=eB/m$ and $\beta= e^2/4$. In the rotating frame,  all the quadratic centrifugal terms have been included into the definition of an arbitrary Lagrange multiplier which does not contribute to the dynamics. We now define $\hat{\omega}_z^2=\omega^2_z+2\beta/(ma^3)$. Let also use  a new coordinate frame $(\xi,\zeta)$  defined by a rotation of angle $\omega_{\lambda} t$. In this case, the rotating wall quadrupole perturbation becomes
\begin{eqnarray}
\nonumber - m \omega_z^2/2 (x^2-y^2)\lambda\cos(2\omega_{\lambda}t) + \nonumber m\omega_z^2 (xy)\lambda \sin(2\omega_{\lambda}t)\rightarrow \\
\nonumber - m\omega_z^2/2(\xi^2-\zeta^2)\lambda,
\end{eqnarray}
which lead us to obtain the Euler-Lagrange equations (we follow almost exactly ~Ref.~$12$),
\begin{widetext}
\begin{eqnarray}
\label{autonomousystem}
 \nonumber \ddot{\xi}-(\Omega-2\omega_{\lambda})\dot{\zeta}+\{\omega_{\lambda}(\Omega-\omega_{\lambda})-(\hat{\omega}_z^2/2-\lambda\omega_z^2) \}\xi=0 \\
 \ddot{\zeta}+(\Omega-2\omega_{\lambda})\dot{\xi}+\{\omega_{\lambda}(\Omega-\omega_{\lambda})-(\hat{\omega}_z^2/2+\lambda\omega_z^2) \}\zeta=0.
 \end{eqnarray}
 \end{widetext}
 Their solutions are
 \begin{widetext}
 \begin{eqnarray}
 \label{BollingerSolutions}
 \nonumber \xi=A_+\cos(\lambda_+ t)+A_-\cos(\lambda_- t),\;\;\;  \zeta=c_+ A_+\sin(\lambda_+ t)+c_- A_-\sin(\lambda_- t),
 \end{eqnarray}
 \end{widetext}
where $A_\pm$ are constants. The frequencies $\lambda_{\pm}$ and the constants $c_{\pm}$ are given by
\begin{widetext}
\begin{eqnarray}\label{lambdapm}
\nonumber \lambda_{\pm}=\frac{1}{2}\{\Omega^2 -2\hat{\omega}_z^2+ (\Omega-2\omega_{\lambda})^2\pm \sqrt{4\hat{\omega}_z^2\omega_z^2\lambda+(\Omega^2-2\hat{\omega}_z^2)(\Omega-2\omega_{\lambda})^2}\},
\end{eqnarray}
\begin{equation}\label{Coefficients}
  c_{\pm}=\frac{\lambda_{\pm}^2-\omega_{\lambda}(\Omega-\omega_{\lambda})+
  \frac{1}{2}(\hat{\omega}_z^2-2\lambda\omega_z^2)}{\lambda_{\pm}(\Omega-2\omega_{\lambda})}.
\end{equation}
\end{widetext}
The system of equations in Equations~(\ref{autonomousystem}) is satisfied for each ion in the trap. Recall that,  owing to the symmetry of the problem, any pair of statistically identical charged density clumps in a Coulomb lattice, will also obtain the same solutions at the corresponding equilibrium positions. In general,  the motion of  this system  is unstable in three dimensions. The more stable configurations should be those with the charged density clumps oscillating  in the $x-y$ plane. As shown in Ref.~$22$, it is consistent with the rotating quadrupolar frequency {\it stroboscopic} election
\begin{eqnarray}
\label{Dubinelection}
 \nonumber \omega_{\lambda}\rightarrow \omega_-, \;\;
  \hat{\omega}_z^2\rightarrow 2(\Omega-\omega_{\lambda})\omega_{\lambda}
\end{eqnarray}
where $\omega_-$ is the trap magnetron frequency.
We will simplify the formulas introducing the trap angle \[\sin\Phi=\sqrt{2}\omega_z/\Omega. \]
In terms of the angle $\Phi$ the magnetron frequency is simply $\omega_-=\Omega\sin^2\frac{\Phi}{2}$ while the cyclotron frequency becomes $\omega_+=\Omega\cos^2\frac{\Phi}{2}$.
Interestingly, in the limit of a thin disk of ions, the equilibrium radius $a$ must be
\begin{equation}\label{a3}
a\rightarrow (\frac{\beta/m}{\omega_{\lambda}\Omega-\omega_{\lambda}^2-\omega_z^2/2})^{1/3}.
\end{equation}
Which can take any limit, i.e., it remains undetermined by the perturbed Penning trap model.
On the other hand, whenever Equations~(\ref{Dubinelection})  are satisfied, the terms depending on $\lambda_-$ in Equation~(\ref{BollingerSolutions} )become irrelevant since, in this case
\begin{eqnarray}\label{lambdaminus}
\nonumber \lambda_-\rightarrow 0, \;\;\;
\lambda_+\rightarrow \omega_+-\omega_-=\Omega\cos\Phi ,
\end{eqnarray}
which leads to select $A_-=0$. Moreover, a rotation of  angle $\lambda_+ t $ leads to the ion center of mass coordinate frame $(x_{\lambda_+}(t),y_{\lambda_+}(t))$. In this system,  when the trap angle $\Phi\ll \pi/2$, the positions $y_{\lambda_+}(t)_{1,2}\rightarrow 0$ and  every two ions lay in opposed positions at a distance $x_{\lambda_+}(t)_1-x_{\lambda_+}(t)_2\simeq 2a$, while the cyclotron motion remains as a rapid oscillation around those adiabatically quasi-stable positions.

\medskip

{\it Mathieu resonances.}
Given that the quadratic Lagrangian Equation~(\ref{L2}) uses only the first two terms in the series of the nonlinear interaction potential energy, when the trap angle $\Phi\ll \pi/2$, the positions should only be stable during a very short period of time of the order of $1/\lambda_+\sim 1/\Omega$. To cope with this difficulty, one should, in general, consider a new dynamic degree of freedom: the polar radial coordinate $\varrho$.  Consequently, one should  replace the constant  $a$ by a function of time $\varrho(t)$, which, indeed, ought to  evolve adiabatically in a period of the order of $1/\omega_{\lambda}\gg 1/\lambda_+$.  Therefore, for each of the individual charges the effective Lagrangian for this new dynamic degree of freedom becomes

\begin{equation}\label{Effective_Mathieu_Lagrangian}
  \mathbf{L}_{\omega_{\lambda}} (\varrho,\dot{\varrho})= \frac{1}{2}m\dot{\varrho}^2+\frac{1}{4}m\hat{\omega}_z^2\varrho^2-\frac{1}{2}m\omega_z^2 \varrho^2\lambda\cos(2\omega_{\lambda} t)
\end{equation}
and the effective time periodic Hamiltonian  becomes
\begin{equation}\label{Effective_Periodic_Hamiltonian}
  H_{\lambda} (\varrho,\dot{\varrho},t)= \frac{\hat{p}^2}{2m}-\frac{1}{4}m\hat{\omega}_z^2\varrho^2+\frac{1}{2}m\omega_z^2 \varrho^2\lambda\cos(2\omega_{\lambda} t)
\end{equation}
The two ions rotate with an angular frequency $\omega_{\lambda}$.  $\varrho(t)$ is the solution of the Mathieu equation,
 \begin{equation}\label{Mathieu}
   \frac{d^2}{d\tau^2}\varrho-\{\mu-2\phi \cos 2\tau \}\varrho=0.
 \end{equation}
In Equation~(\ref{Mathieu}) $\tau=\omega_{\lambda} t$, $\mu=\cot^2\frac{\Phi}{2}$, and $\phi=\lambda\mu$. The solutions are written in terms of the oscillatory Mathieu cosine functions
\begin{equation}\label{Mathieu_Cos_Theta}
  \varrho(\tau)=  a   C_e(-\mu,-\phi,\tau)/C_e(-\mu,-\phi,0).
\end{equation}
Nonetheless, there would only be  periodic stable solution within a very narrow parametric region $\phi(\mu)$ (see Ref.~$23$ for reviewing the entire parametric map); these have $\pi$ period for the variable $\tau$.  When $\Phi\rightarrow 0$, the first order parametric stability constraint is
\begin{equation}\label{Periodic-Condition}
\phi(\mu)\sim \mu/2+o(\sqrt{\mu}).
\end{equation}
This largely oscillatory behaviour corresponds to a radial breathing collective motion of the  Coulomb lattice, i.e., a  new  degree of freedom. Finally, if Equation~(\ref{Periodic-Condition}) is satisfied, the Euler-Lagrange equation Equation~(\ref{Mathieu}) reads
\begin{equation}\label{Mathieu2}
   \frac{d^2}{d\tau^2}\varrho-\mu \varrho[2\sin^2\tau +o(1/\sqrt{\mu})]\simeq 0.
\end{equation}
Since the solutions of Equation~(\ref{Mathieu2}) are necessarily periodic, in order to physically understand the motion of the ion in the Penning trap, an average  of the periodic term will be now obtained (assuming that  $\langle2\sin^2\tau\rangle= 1$ during  many loops of its orbit). The average motion is identical to that of an  inverted harmonic oscillator for $\langle \varrho(\tau)\rangle$. The orbits should be  restricted between a maximum and a minimum $\varrho(t)$. Far from the turning points at $\tau=0$ and $\tau=\pi$,  one has
\begin{equation}\label{Mathieu_mu_limit}
  \langle \frac{d^2}{d\tau^2}\varrho-\mu \varrho \rangle = 0  .
\end{equation}
In that limit the  Lagrangian becomes
\begin{equation}\label{Effective_Mathieu_Lagrangian_lowLimit}
  \mathbf{L} \rightarrow \langle\frac{1}{2}m\dot{\varrho}^2+\frac{m\omega_z^2\varrho^2}{4}\rangle
\end{equation}
and average Hamiltonian reads
\begin{equation}\label{Hamiltonian_Inverted_harmonic_Oscillator}
  H_0 \rightarrow \langle \frac{\hat{p}^2}{2m}-m\frac{\omega_z^2}{4}\varrho^2\rangle,
\end{equation}
which coincides with the postulated Hamiltonian of the factorisation function.

\medskip

%

\end{document}